# Facile residue analysis of recent and prehistoric cook-stones using handheld Raman spectrometry

Laura Short,[1] Alston V. Thoms[1], Bin Cao,[2,3] Alexander M. Sinyukov,[2] Amitabh Joshi,[4] Rob Scully,[2] Virgil Sanders,[2] and Dmitri V. Voronine[2,4]

[1]*Department of Anthropology, Texas A&M University, College Station, TX*
[2]*Institute for Quantum Science and Engineering, Department of Physics and Astronomy, Texas A&M University, College Station, TX*
[3]*Xi'an Jiaotong University, Xi'an, China 710049*
[4]*Department of Physics, Baylor University, Waco, TX*

We performed food residue analysis of cook-stones from experimental and prehistoric earth ovens using a handheld Raman spectrometry. Progress in modern optical technology provides a facile means of rapid non-destructive identification of residue artifacts from archaeological sites. For this study spectral signatures were obtained on sotol (*Dasylirion* spp.) experimentally baked in an earth oven as well as sotol residue on an experimentally used processing tool. Inulin was the major residue component. The portable handheld Raman spectrometer also detected traces of inulin on boiling stones used to boil commercially obtained inulin. The Raman spectra of inulin and sotol may be useful as signatures of wild plant residues in archaeology. Spectroscopic analysis of millennia-old cook-stones from prehistoric archaeological sites in Fort Hood, TX revealed the presence of residues whose further identification requires improvement of current optical methods.

**Introduction**

The first analysis of archaeological food residue occurred in the 1930s, when Johannes Grüss used basic chemical tests to identify black residue on a ceramic vessel as overcooked milk [1]. Since then, residue analysis has been conducted on a wide variety of substances including perfumes, cosmetics, beeswax, resins, tar, pitches, proteins and lipids in soils, pigments, ink, and paint [2-5]. Food residue studies generally analyze lipids, proteins, DNA, and other characteristic compounds of residues absorbed by pottery. A wide range of techniques is used including chromatography, gas spectrometry, elemental analysis, optical and resonance spectroscopy, stable isotope analysis, X-ray diffraction and immunological techniques [6].

Studies of food residue have been most successful with pottery, likely because the porous nature of the pottery enables substances to become easily absorbed and trapped. There also have been successful protein and lipid analyses of residues on the surface of grinding implements and flaked tools [6]. In both cases, blind tests using modern laboratory-created artifacts have shown that these methods are in need of further development and utilization of multiple lines of evidence [7-9].

Raman spectroscopy for archaeological analysis has focused on paints and pigments, resins and pitch, and plaster-like materials [4, 6]. It can be used to identify both organic and inorganic substances and has gained popularity due to its non-destructive nature. However, fluorescence background may limit the sensitivity and archaeological materials may undergo taphonomic processes that make matches to modern reference samples difficult [10]. Additionally, until recently, Raman analysis has been laboratory oriented.

Various types of Raman instruments have been developed and optimized for different purposes. A class of miniaturized portable Raman spectrometers is now available for rapid *in situ* experiments such as airport screening, forensics, art authenticity verification, etc. Handheld Raman spectrometers can be used by a single operator in diverse challenging environments and may be particularly useful in archaeology, especially in situations when artifacts cannot be easily moved to the laboratory or when objects are too large for a microscope. Several applications of portable spectrometers to examine the composition of compounds in art such as canvas and rock paintings have been recently reported [11-13].

In this paper, we use a handheld Raman spectrometry to perform trace analysis of food residue from limestone rocks used experimentally as heating elements (i.e., cook-stones) in actualistically constructed and used earth ovens. We also analyzed cook-stones recovered from prehistoric earth ovens at archaeological sites in Fort (Ft.) Hood, TX.

**Background: Hot-Rock Cooking Techniques**

Cook-stone technology, the use of heated rocks for cooking, is roughly 30,000 years old, and has occurred worldwide. Techniques include using heated stones as griddles in open hearths, as heating elements in closed earth ovens and steaming pits, and as the heating element for boiling. Its appearance in the archaeological record has been related to population packing that required people to put more effort into procuring more food from the same area of land. This technology requires more energy input than hot coal cooking, because stones and green-plant packing material have to be collected in addition to the firewood; however, it is more fuel efficient because the stones retain heat long after the coals cool [14, 15].

In their most essential form, earth ovens consist of a pit in which heated stones are used to cook food. Generally speaking, food may or may not be wrapped into packages, but is always insulated from the stones with green plant material. Earth ovens are ideal for cooking foods that require a long cooking time. Ethnographic evidence shows that many groups around the world cooked meat, fish and shellfish in earth ovens. Most archaeological evidence indicates that pre-Columbian (i.e. prehistoric) North Americans living in temperate environs most commonly cooked plants in earth ovens. In the eastern half of Texas, geophytes, especially bulbs of eastern camas (*Camassia scilloides*), wild onion (*Allium* spp.), and false garlic (*Nothoscordum bivalve*) were baked in earth ovens as early as 8-9,000 years ago. Geophytes are perennial plants that winter-over and propagate via underground buds (e.g. bulbs, corms, tubers, rhizomes). In the western half of Texas desert succulents were commonly baked in ovens, including lechuguilla (*Agave lechuguilla*), sotol (*Dasylirion* spp.), and prickly pear (*Opuntia* spp.) [14, 15]. For the most part, knowledge about what was baked in earth ovens comes from ethnographic evidence and occasionally carbonized plant remains from archaeological remains of earth ovens.

Cook-stones were also used to boil water, in a process known as stone boiling [14, 15]. In this case, stones heated in an open fire to about 500 ºC were removed using tongs, quickly rinsed in water, and dropped into a vessel containing liquid and food.  As the stones cooled, they were removed and hot ones were added until the food was adequately boiled.  This method boils liquids in bark, wooden, or hide containers more quickly than direct heating methods, and it does not require heat-resistant materials (e.g., ceramic and metal) as do direct heating methods.  Stone boiling was used for a wide variety of cooking applications, creating soups, stews, porridge, and rendering fat.  Many foods were cooked by stone boiling - nuts and seeds, geophytes, meat, and fish.  Nuts and animal parts were both used to render fat.  Since stone boiling does not usually result in charred materials, at this point most knowledge of what was cooked by this method is based on ethnographic evidence [15].

Starch granule and other residue analyses are now being used to identify plant-food microfossils in cooking stones, albeit with  mixed results [16].  Raman spectroscopy also provides the potential to identify what was in direct contact with the cook-stones used in boiling as well as minute food remains adhering to rocks used as heating elements in earth ovens.

**Background: Plant Carbohydrates**

Plant carbohydrates include simple sugars and alcohols, storage polysaccharides and structural polysaccharides.  Simple sugars such as glucose and fructose make up the sweetness we taste in fresh fruits and vegetables.  Storage polysaccharides such as starch and fructans are used to store energy.  Structural polysaccharides such as cellulose and pectin are the components of cell walls known as dietary fiber [17].

A specific storage carbohydrate, inulin, is associated with earth-oven  baking[15]. Inulin is concentrated in the edible underground storage organs (bulbs, tubers, etc.) of some geophytes including many plants in the lily family, such as onion and garlic (*Allium* spp.) and camas (*Camassia* spp.), and many plants in the aster family, including chicory (*Cichorium intybus*), jerusalem artichoke (*Helianthus tuberosus*), and dandelion (*Taraxacum* spp.), as well in the hearts of succulents such as sotol (*Dasylirion* spp.) and agave (*Agave* spp.).  The simpler the carbohydrate, the easier it is for humans to digest and utilize the sugar – complex carbohydrates such as starch and inulin must undergo hydrolysis to be readily digestible.  Raw inulin provides energy via digestion by gut flora (which is why it is known as a prebiotic), but inulin breaks down into simpler sugars fructose and glucose when cooked over a long period of time.  Earth ovens, which are capable of generating and maintaining sufficient heat for 72 hours, are ideal for the kind of extended cooking required to break down inulin and thereby render it more readily digestible [17].

**Hot-Rock Cook-Off: Experiment and Analysis**

The Hot-Rock Cook Off (HRCO) is an actualistic experimental archaeological cooking event, where cooking methods utilizing hot rocks are recreated based on archaeological and ethnographic data. Earth oven-cooking is the focus of the event, though stone boiling and grilling are included. Predominantly an academic venture by anthropology students at Texas A&M and Texas State Universities, it is open to the public and includes other educational activities and information. Each year representatives of Native American groups from the region attend and participate in the event. These experiments can be thought of as an attempt to replicate archaeological signatures of earth ovens found throughout Texas and elsewhere around the world. Figs. 1A and 1B depict earth ovens used during the HRCO event in San Marcos, TX in November 2012. To replicate prehistoric cooking techniques, sotol was baked for approximately 48 hours using heated limestone rocks (Figs. 2A and 2B).

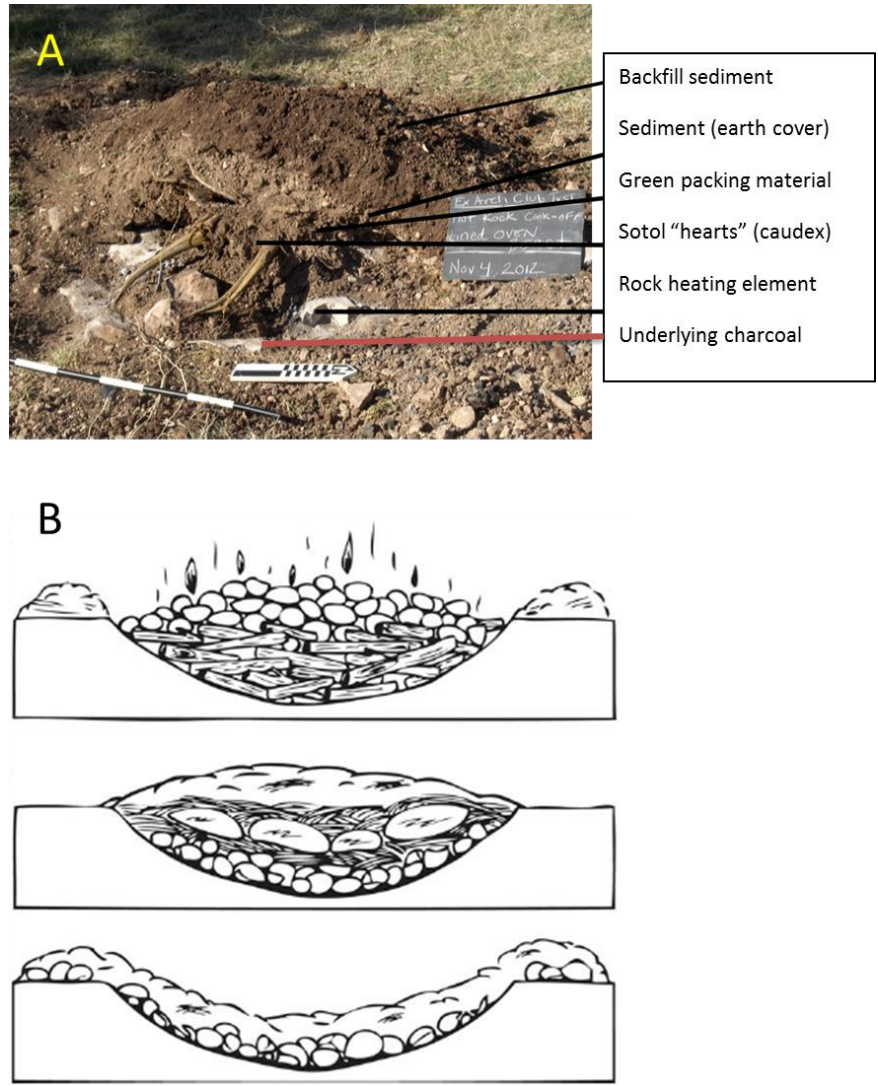

Figure 1. (A) Photograph of a partially uncovered 48-hour earth oven from the HRCO field experiments in San Marcos, TX that was used to bake sotol. (B) Schematic illustration of construction and use of a typical earth oven (adapted from Thoms, A. V. J. Anthropol. Archaeology 27, 443 (2008)): (B, top) fire is built in a pit overlain by a layer of rocks; (B, middle) when the fire burns completely, red-hot rocks are covered with green packing material, food packs, more packing material, and covered with earth; and (B, bottom) remains of the oven after the food is removed and the oven is abandoned.

We measured the Raman spectra using the 'First Guard' handheld Raman spectrometer from the Rigaku Corporation, which has a 1064 nm laser, a spectral resolution of ~20 cm$^{-1}$, and a detection range from 200 to 2000 $cm^{-1}$. The 1064 nm wavelength provides advantages of *in situ* investigation and a significant suppression of fluorescence background. This push-button device is most convenient for field experiments that do not require sample preparation. It is therefore especially suitable for non-destructive efficient exploration of prehistoric archaeological sites.

At the HRCO, a stone tool was used to scrape the baked sotol and make it into cakes more suitable for eating, as is documented ethnographically and likely occurred in the distant past as well (Fig. 2C) [18]. The handheld Raman spectrometer was used to examine the visible residue that remained on the scraper. The laser beam was focused on the stone surface at 400 mW laser power with 3 s exposure time and an average of three shots. The results are shown in Fig. 2D. Both the surface of the scraper (black) and the spectra from fresh sotol (red) have a well-resolved peak at 1453 $cm^{-1}$ which is absent in a clean stone washed with tap water. There are also other peaks around 800 and 1100 cm$^{-1}$ which are weak. These peaks confirm the presence of sotol on the surface of the scraper. The signal intensity varied depending on the position on the scraper. The results imply that the key issue in detecting residues on artifacts is to find a hotspot where some residue adheres to the surface or in cracks and crevices. That several places on a given artifact can be sampled in a short timeframe indicates the practicality of handheld Raman spectrometry in field and laboratory archaeology.

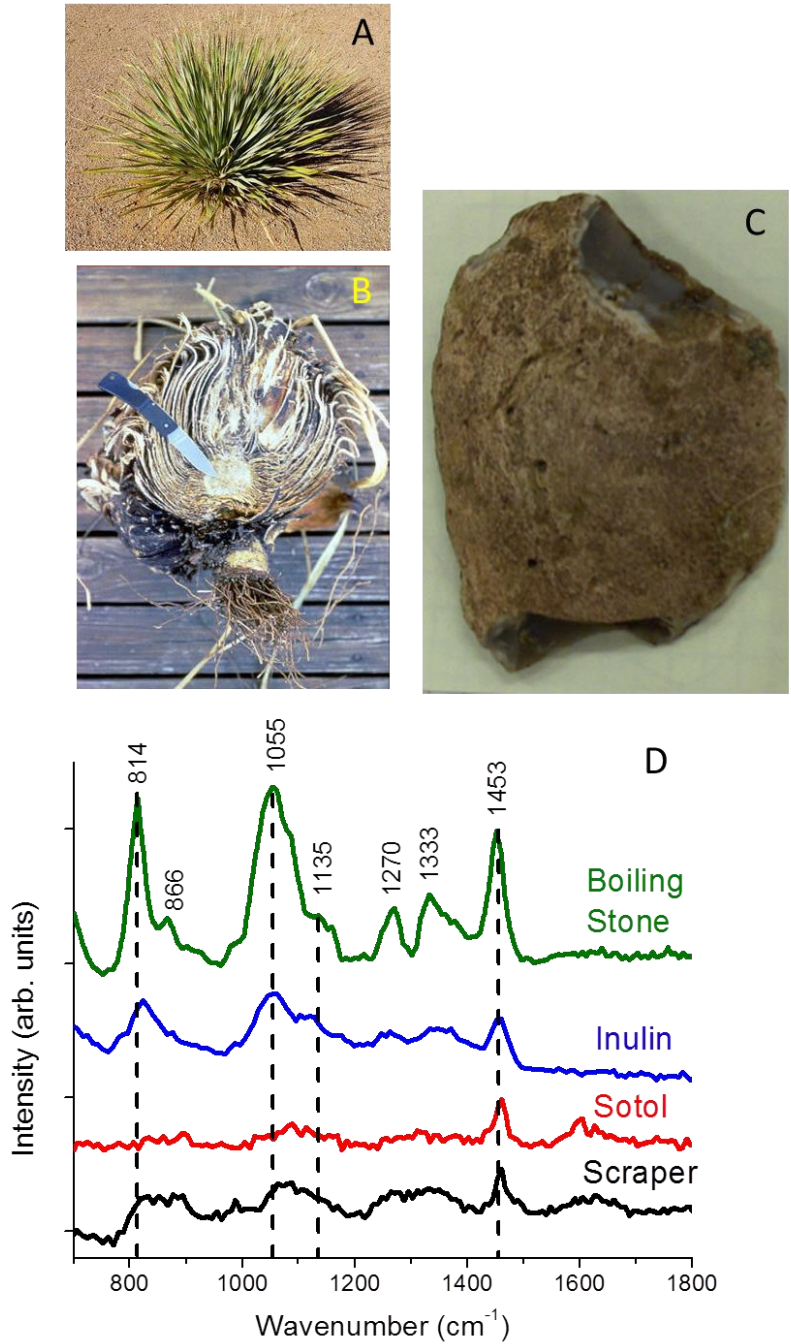

Figure 2. (A) Sotol. (B) A desiccated sotol "heart" sliced with a saw to show the internal structure of the plant. The knife points to the edible central stem from which the leaves grow, something like an artichoke. (Courtesy of Phil Dering) (C) Stone tool used to scrape cooked sotol at the HRCO site. (D) Raman spectra of cooked sotol (red) and sotol residue on the scraper (black), compared with uncooked inulin (blue) and cooked inulin on boiling stones (green). Similar spectral signatures were found in all samples.

We compared the spectra of the raw sotol and baked sotol on the surface of the scraper to the spectra of inulin. We also used handheld Raman spectrometry for residue analysis of limestone fragments used to boil chicory root inulin powder purchased from a local grocery store. The limestone was purchased from a local garden center. About 5 grams of inulin were boiled with several stones for an hour.

Fig. 2D shows a comparison of Raman spectra of raw inulin (blue) with cooked inulin on the surface of boiling stones (green), and with raw sotol (red) and baked sotol on the surface of the scraper (black). The obtained spectra of inulin are in agreement with previous reports [19, 20]. Spectra of cooked inulin on boiling stones reveal clear signatures of inulin. Sotol and inulin have similar spectra. Therefore inulin is a major component in Raman spectra of sotol. This confirms the potential of handheld Raman spectrometry for archaeological food residue analysis on boiling stones.

**Prehistoric cook-stones:  Methods and Analysis**

We examined two cook-stones, commonly known as fire-cracked rocks (FCR) from two ancient earth ovens.  These FCR were among many such cook-stones constituting the heating element of earth ovens excavated at Ft. Hood, TX. Figs. 3B-3D and 3F-3H show photographs of different sides of *stones 1* and *2*, respectively. *Stone 1,* from site 41CV1553, dates to approximately 350-650 AD. *Stone 2,* from the site 41CV594, dates to approximately 2,500-500 BC. Raman spectra from the surface of *stones 1* and *2* are shown in Figs. 3A and 3E, respectively. As described above, the spectrometer was put against the surface of the cook-stones to obtain the spectra, and different spots were selected. A small piece cut from *stone 1* was thoroughly cleaned for comparison (Fig. 3J). The corresponding Raman spectrum is shown in Fig. 3I. The Raman spectra in Figs. 3A, 3E and 3I show similar patterns. Both stones showed Raman peaks around 988, 1085, and 1170 cm$^{-1}$. The same peaks were also found on the piece of *stone 1* that was cleaned with tap water. Therefore, they are assigned to the stone itself. However, the spectra of several spots on the uncleaned cook-stones showed broadening of the 1085 cm$^{-1}$peak. This broadening was not observed on the cleaned cook stone and is attributed to the presence of residues.

Fig. 3 shows that the spectra 1 and 3 in (A) and the spectrum (I) of the section of the cook-stone cleaned by tap water have a narrower width at 1085 $cm^{-1}$ compared to the spectra from the surface of the stones. It is possible that the observed broadening of the peak at 1085 $cm^{-1}$ is due to organic food residues such as carbohydrates. Inulin is present in many wild plants found in the vicinity of the sites, especially onion and camas, both of which have been recovered as charred macrobotanical fragments from remains of ancient oven at Fort Hood [21]. However, other inulin spectral peaks such as the 1453 $cm^{-1}$ peak were not resolved due to low signal-to-noise ratio. This finding suggests the possibility of identifying organics, including residue of food eaten a thousand or more years ago, using handheld Raman spectrometry. Assessment of this working hypothesis—broadening of the peak at 1085 $cm^{-1}$ is due to organic food residues — requires improvement of the signal-to-noise, spectral resolution and extension of the detection spectral range.

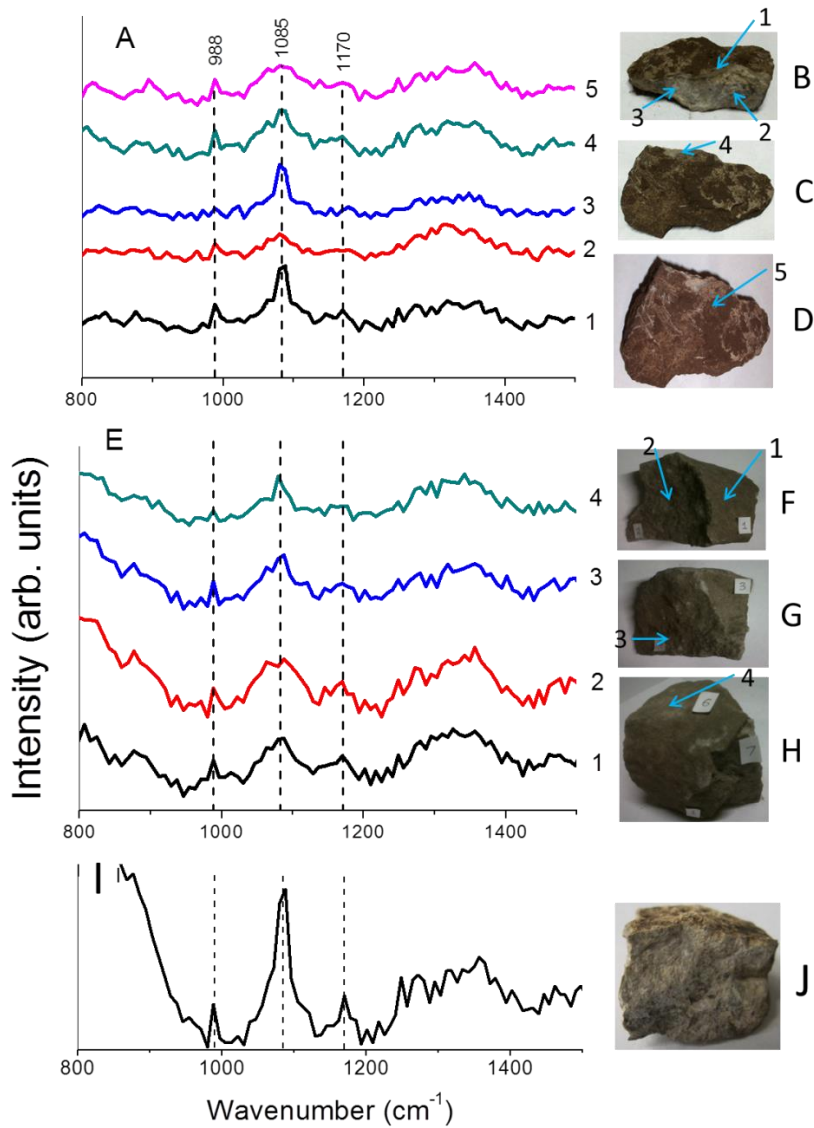

Figure 3. (A) and (E) are Raman spectra of two different stones from the prehistoric archaeological sites in Ft. Hood, labeled *stone 1* and *stone 2*, respectively. (B) - (D) and (F) - (H) are photographs of different sides of *stones 1* and *2*, respectively. (B) is a split cross-section of *stone 1* with the corresponding spectra *1 - 3* in (A). (I) Raman spectrum of a cracked piece of *stone 1 (J)* after cleaning with tap water. Arrows indicate spatial positions on the cook-stones that correspond to the spectra. The cook-stone sizes vary in the range 3 – 15 cm.

**Comparison of the Portable and Lab-based Raman Instruments**

We compared the performance of the portable handheld Raman spectrometer with the state-of-the-art lab-based Raman microscope. The latter was a confocal Raman microscope (Nanonics Imaging, Ltd) with an electric-cooled CCD detector (-70 °C) and iHR550 spectrometer (Horiba), and 180° backscattering detection. The excitation source was a 785 nm cw laser with up to 30 mW power at the sample with a 10x objective. The spectral resolution was better than 0.7 cm$^{-1}$. To perform the comparison of the two instruments we purchased two reference materials, inulin from chicory root and cellulose acetate, from Sigma-Aldrich, Inc. Both of these materials may be present as food residues at archeological sites. The corresponding Raman spectra of inulin and cellulose are shown in Figs. 4A and 4B, respectively. The comparison of the spectra measured using the portable (red) and lab-based (blue) instruments shows that both instruments provide essentially the same information. The portable instrument has lower spectral resolution but is still able to detect most of the spectral lines. This demonstrates that the portable Raman instrument may be used for residue analysis.

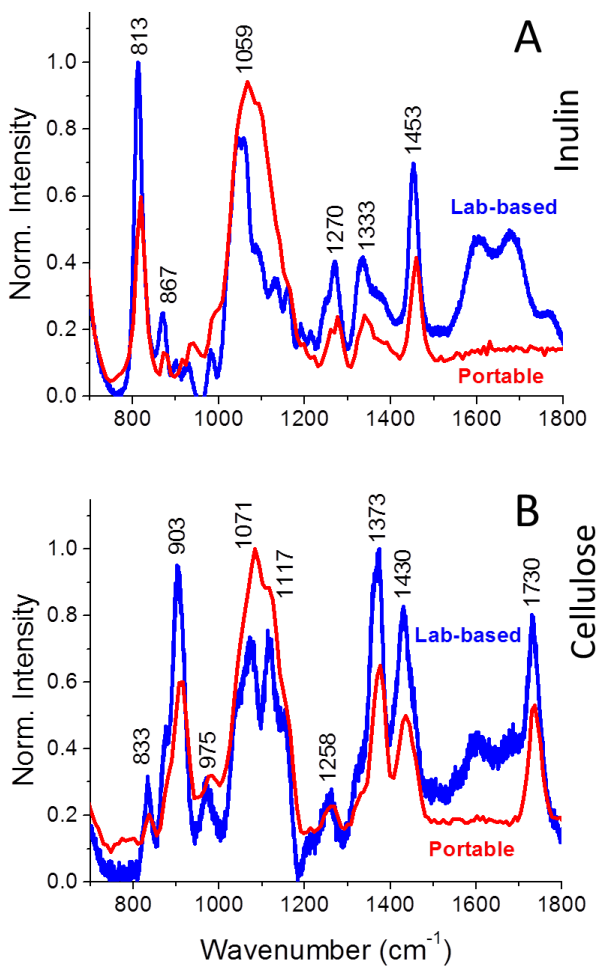

Figure 4. Raman spectra of inulin (A) and cellulose (B) purchased from Sigma-Aldrich, Inc measured with a lab-based (blue) and portable (red) instruments. Similar spectral signatures obtained with both devices demonstrate that a portable instrument can be used in archeological field experiments.

Figure 5 shows a comparison of Raman spectra of inulin from a grocery store (blue) to the chemical grade inulin from Sigma-Aldrich (red) measured using the handheld spectrometer. Similar results are obtained. This shows that the portable Raman spectrometer can detect inulin from various sources.

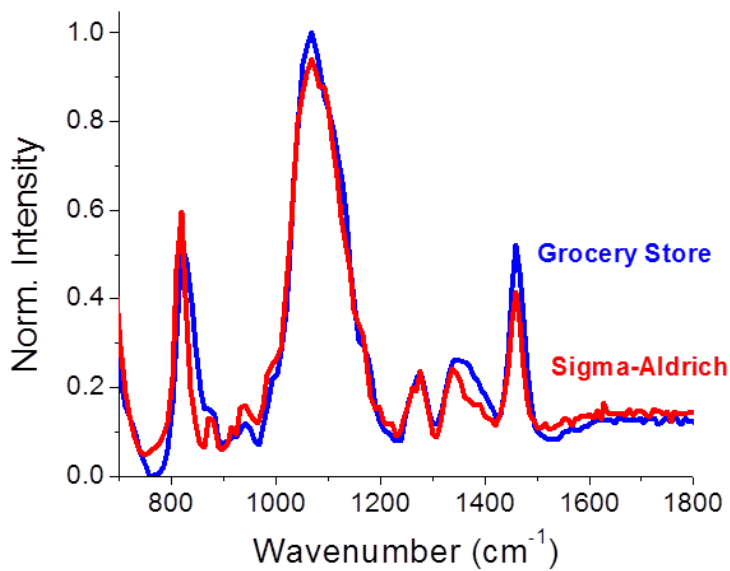

Figure 5. Comparison of the Raman spectra of inulin from Sigma-Aldrich, Inc (red) and from a grocery store (blue) obtained using a handheld spectrometer. Similar spectral signatures in both cases are observed.

Band assignment was performed based on previous Raman and FTIR studies of fructose, inulin and cellulose. Comparison of the bands of inulin and cellulose in Table 1 shows that these two different carbohydrates can be distinguished using portable Raman spectroscopy. For example, the CH$_2$-OH bending and deformation bands at 1333 and 1453 cm$^{-1}$ in inulin are suppressed and shifted in cellulose.

| Inulin (cm$^{-1}$) | Cellulose (cm$^{-1}$) | Band Assignment |
|---|---|---|
| 813 s | - | CC stretching |
| - | 833 w | CCC, COC, OCC, OCO skeletal bending |
| 867 w | - | COC bending |
| - | 903 s | HCC, HCO bending |
| - | 975 w | HCH bending |
| 1059 s | - | COC stretching and ring deformations |
| - | 1071 s | COC stretching symmetric |
| - | 1117 s | |
| - | 1258 w | HCH (twisting), HCC, HOC, COH (rocking) bending |
| 1270 s | - | CH bending |
| 1333 s | - | CH$_2$-OH bending and deformations symmetric |
| - | 1373 s | HCH, HCC, HOC, COH bending |
| - | 1430 s | HCH asymmetric |
| 1453 s | - | CH$_2$-OH bending and deformations asymmetric |
| - | 1730 s | C=O stretching |

Table 1. Summary of Raman shifts and band assignments of inulin and cellulose from Sigma Aldrich (s – strong, w - weak). The band assignment was based on previously reported Raman and FTIR spectra of fructose, inulin and cellulose.

Inulin has also a broad band around ~2900 cm$^{-1}$ (not shown), which lies outside of the available range of the handheld Raman spectrometer (from 200 to 2000 cm$^{-1}$). This band, however, cannot be used for inulin identification because it is present in all carbohydrates. The available "finger-print" spectral range is sufficient to identify inulin at archeological sites using portable measurements (Figure 4). Further analysis in a broader range can be later performed using laboratory-based instruments.

## Conclusions

We demonstrated the use of handheld Raman spectrometry for facile trace analysis of inulin in actualistic experiments and its potential application at prehistoric archaeological sites. We detected spectroscopic features of inulin in the Raman spectra of sotol, which is a potential residue source in prehistoric earth ovens. Future exploration of archaeological samples using handheld Raman spectrometers is anticipated. Given that food residue is most likely to be preserved in the cracks and crevices of ancient, well weathered cook-stones and tools [16, 22], we conclude that portable handheld Raman microscopy should focus on these places on a given stone [10].

Coherent anti-Stokes Raman scattering (CARS) spectroscopy has been recently used for the investigation of the molecular composition of gas residues in cracks of translucent materials [23, 24]. CARS can be also used for the archaeological food residue analysis. Another possible future direction of improving handheld Raman spectrometry is by increasing the sensitivity via surface enhancement [25, 26]. Surface-enhanced Raman scattering (SERS) micro-spectroscopy has been used for the detection of nucleotide traces in pyroxene rocks as imitation of *in situ* search for life traces on Mars [27]. It may also be possible to adapt combinations of these techniques to the *in situ* food residue analysis and to develop portable surface-enhanced CARS (SECARS) [28, 29] and FAST-CARS [30, 31] spectrometers. Developing portable handheld CARS, SERS and SECARS spectrometers may bring many future advantages in the field.

## Acknowledgements

The United States Army Fort Hood provided some of the funds for this research, and special thanks go to Richard Jones, director of Fort Hood's cultural resource management program, for supporting this effort. Jones and Doug Boyd (Prewitt and Associates, Inc. in Austin, Texas) facilitated access to the specimens used in this study. We gratefully acknowledge Professor Marlan O. Scully for helpful discussions. Drs. Stephen Black and Jon Lohse, of the Department of Anthropology and the Center for Archaeological Study, respectfully at Texas State University provided logistical support. Elizabeth Vilchez, undergraduate student in the Department of Anthropology, assisted in field and lab work for this study. Additional funding and other support was provided by the Department of Anthropology and Archaeological Ecology Lab at Texas A&M University.